\documentclass[conference]{IEEEtran}
\IEEEoverridecommandlockouts
\usepackage{cite}
\usepackage{amsmath,amssymb,amsfonts}
\usepackage{algorithmic}
\usepackage{graphicx}
\usepackage{textcomp}
\usepackage{xcolor}
\usepackage{hyperref}
\usepackage{balance} 
\usepackage{lipsum} 
\usepackage{multirow}
\usepackage{tikz}

\newcommand\copyrighttext{%
  \footnotesize \textcopyright \the\year{} IEEE. Personal use of this material is permitted. Permission from IEEE must be obtained for all other uses, including reprinting/republishing this material for advertising or promotional purposes, collecting new collected works for resale or redistribution to servers or lists, or reuse of any copyrighted component of this work in other works.}

\newcommand\copyrightnotice{%
\begin{tikzpicture}[remember picture,overlay]
\node[anchor=south,yshift=10pt] at (current page.south) {\fbox{\parbox{\dimexpr0.75\textwidth-\fboxsep-\fboxrule\relax}{\copyrighttext}}};
\end{tikzpicture}%
}

\def\BibTeX{{\rm B\kern-.05em{\sc i\kern-.025em b}\kern-.08em
    T\kern-.1667em\lower.7ex\hbox{E}\kern-.125emX}}
    
\begin{document}

\title{SynthBA: Reliable Brain Age Estimation Across\\Multiple MRI Sequences and Resolutions}

\makeatletter
\newcommand{\linebreakand}{%
  \end{@IEEEauthorhalign}
  \hfill\mbox{}\par
  \mbox{}\hfill\begin{@IEEEauthorhalign}
}
\makeatother

\author{\IEEEauthorblockN{1\textsuperscript{st} Lemuel Puglisi}
\IEEEauthorblockA{\textit{Dept. of Math and Computer Science} \\
\textit{University of Catania}\\
Catania, Italy \\
lemuel.puglisi@phd.unict.it}
\and
\IEEEauthorblockN{2\textsuperscript{nd} Alessia Rondinella}
\IEEEauthorblockA{\textit{Dept. of Math and Computer Science} \\
\textit{University of Catania}\\
Catania, Italy\\
alessia.rondinella@phd.unict.it}
\and
\IEEEauthorblockN{3\textsuperscript{rd} Linda De Meo}
\IEEEauthorblockA{\textit{Department of Neuroinflammation} \\
\textit{University College London} \\
London, UK \\
e.demeo@ucl.ac.uk}
\linebreakand
\IEEEauthorblockN{4\textsuperscript{th} Francesco Guarnera}
\IEEEauthorblockA{\textit{Dept. of Math and Computer Science} \\
\textit{University of Catania}\\
Catania, Italy\\
francesco.guarnera@unict.it}
\and
\IEEEauthorblockN{5\textsuperscript{th} Sebastiano Battiato}
\IEEEauthorblockA{\textit{Dept. of Math and Computer Science} \\
\textit{University of Catania}\\
Catania, Italy\\
battiato@dmi.unict.it}
\and
\IEEEauthorblockN{6\textsuperscript{th} Daniele Ravì}
\IEEEauthorblockA{\textit{School of Physics, Engineering and CS} \\
\textit{University of Hertfordshire}\\
Hatfield, UK\\
d.ravi@herts.ac.uk}
}

\maketitle

\begin{abstract}
Brain age is a critical measure that reflects the biological ageing process of the brain. The gap between brain age and chronological age, referred to as brain PAD (Predicted Age Difference), has been utilized to investigate neurodegenerative conditions. Brain age can be predicted using MRIs and machine learning techniques. However, existing methods are often sensitive to acquisition-related variabilities, such as differences in acquisition protocols, scanners, MRI sequences, and resolutions, significantly limiting their application in highly heterogeneous clinical settings. In this study, we introduce Synthetic Brain Age (SynthBA), a robust deep-learning model designed for predicting brain age. SynthBA utilizes an advanced domain randomization technique, ensuring effective operation across a wide array of acquisition-related variabilities. To assess the effectiveness and robustness of SynthBA, we evaluate its predictive capabilities on internal and external datasets, encompassing various MRI sequences and resolutions, and compare it with state-of-the-art techniques. Additionally, we calculate the brain PAD in a large cohort of subjects with Alzheimer’s Disease (AD), demonstrating a significant correlation with AD-related measures of cognitive dysfunction. SynthBA holds the potential to facilitate the broader adoption of brain age prediction in clinical settings, where re-training or fine-tuning is often unfeasible. The SynthBA source code and pre-trained models are publicly available at \url{https://github.com/LemuelPuglisi/SynthBA}.
\end{abstract}


\copyrightnotice

\begin{figure*}[h!]
    \centering
    \includegraphics[width=\textwidth]{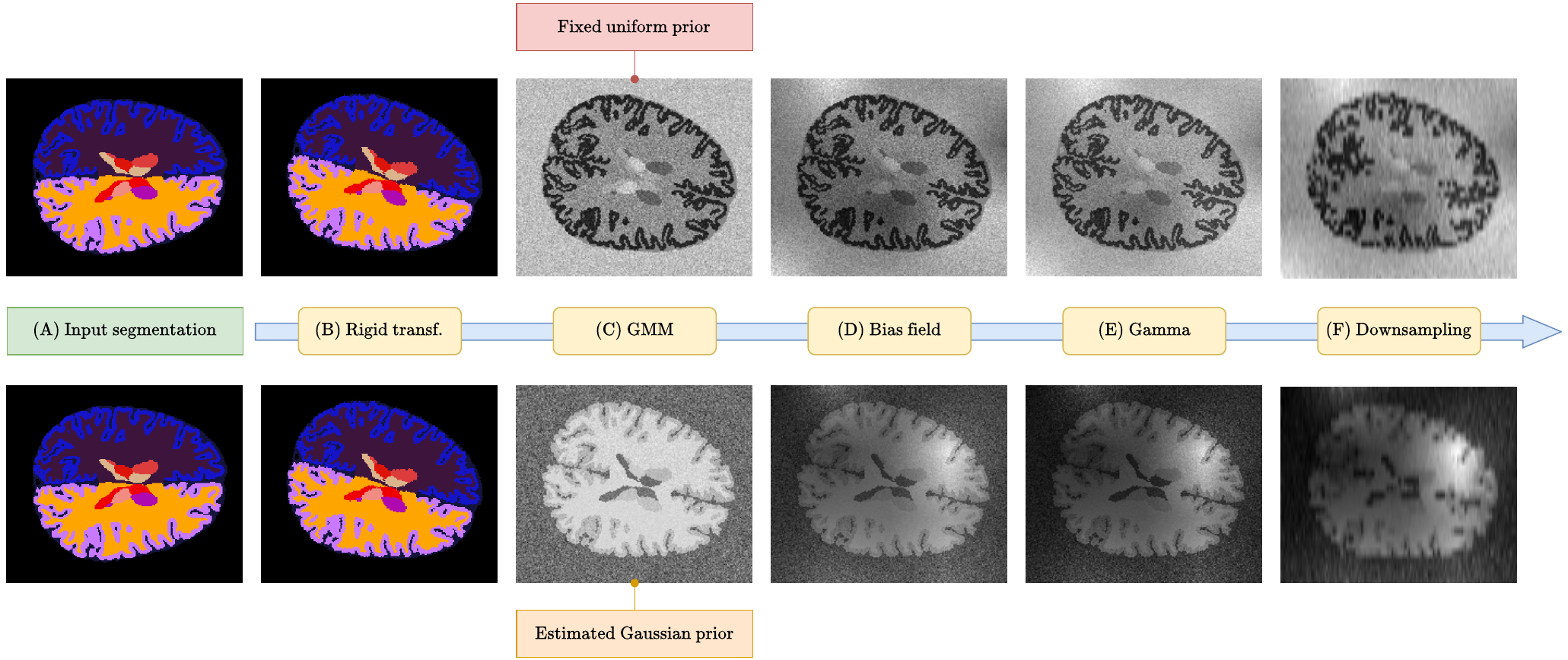}
    \caption{All the steps from the generative process $x \sim \mathcal G (s)$. The upper branch illustrates the generative process when the GMM employs a uniform prior, while the lower branch shows the process when the GMM uses a Gaussian prior estimated from real MRIs.}
    \label{fig:generator}
\end{figure*}

\section{Introduction}\label{sec:introduction}
The biological ageing process of the brain can be accelerated by various health conditions, such as neurodegenerative diseases, and by environmental factors. Recent studies have demonstrated that disentangling the biological age of the brain (referred to as brain age) from the chronological age of the individual aids in predicting disease risk~\cite{cole2018brain}. The primary assumption is that the brain age of healthy individuals closely matches their chronological age. Building on this assumption, researchers have developed machine learning techniques to predict the chronological age of healthy individuals from brain MRI data as a means to estimate their brain age~\cite{cole2017predicting}. When these models are applied to subjects with neurological diseases, an increased discrepancy is observed between their predicted brain age and chronological age~\cite{cole2020longitudinal}. This widening gap, referred to as brain PAD, highlights an accelerated ageing process of the brain in diseased individuals~\cite{huang2021accelerated}. \newline

Deep learning has brought significant advancements to brain age modeling, resulting in models that surpass traditional methods in terms of accuracy. However, their sensitivity to variations in acquisition protocols, scanners, MRI sequences, and resolutions limits their practical use across diverse clinical environments. Despite extensive research done in the medical imaging analysis field to address this issue, such as using domain-specific augmentation techniques~\cite{puglisi2024deepbrainprint} or domain randomization~\cite{tobin2017domain}, only a few studies have been conducted to tackle this challenge in the context of brain age modeling. For example, the authors of~\cite{wood2024optimising} develop five brain age models, each trained on a large dataset of a different MRI sequence: T1-weighted (T1w), T2-weighted (T2w), Fluid-Attenuated Inversion Recovery (FLAIR), Gradient-Recalled Echo (GRE), and Diffusion-Weighted Imaging (DWI). These models are intended to serve as foundational models that can be fine-tuned on a target dataset to improve predictive accuracy compared to training a model ex novo. While the availability of brain age models for various MRI sequences facilitates their adoption, they still require fine-tuning to achieve low errors. Another relevant work is proposed in~\cite{barbano2023contrastive}, which introduces a contrastive regression loss that enhances robustness against site differences~\cite{bayer2022site}, including variations in scanner vendor, head coil, and acquisition parameters. Although this method achieves state-of-the-art performance in the multi-site challenge~\cite{dufumier2022openbhb}, it does not account for higher-level variations, such as different MRI sequences and resolutions. Lastly, \cite{valdes2023toward} proposes a dual-transfer learning strategy to obtain an unbiased, protocol-agnostic brain age model. Nevertheless, their experimental results reveal an unfavorable trade-off between predictive accuracy and generalizability.\newline

Our work stems from the practical necessity of computing brain age using clinical-grade MRIs, regardless of their acquisition settings, without the possibility of re-training an existing model. This scenario is typical in clinical trials, where the selected subjects are often impaired, leaving no healthy cohort available to train or fine-tune existing brain age models. This limitation restricts the applicability and, consequently, the large-scale adoption of brain age assessments. \newline

The ability to generalize across acquisition-related variability is essential in medical imaging applications. Domain randomization~\cite{tobin2017domain} has recently been employed to address this issue in various MRI-related tasks. This approach involves training a deep learning model on synthetic data generated by a simulator with fully randomized parameters. Although this technique can produce unrealistic data, such as MRIs with atypical contrasts, it improves the model's ability to generalize to previously unseen data distributions~\cite{tobin2017domain}. For example, the authors of~\cite{billot2023synthseg} developed a generative model for brain MRIs and trained a brain tissue segmentation model, called SynthSeg, by fully randomizing the generator's parameters. As a result, SynthSeg can adapt to varying contrasts and resolutions. Building on this concept, several robust models have been developed for various MRI-related tasks: SynthStrip~\cite{hoopes2022synthstrip} for skull-stripping, SynthSR~\cite{iglesias2023synthsr} for super-resolution, and SynthMorph~\cite{hoffmann2021synthmorph} for image registration. \newline

Building on the intuition of the seminal work SynthSeg~\cite{billot2023synthseg}, we propose \textbf{SynthBA}, a deep learning-based brain age model robust to acquisition-related variability. We adapted the generative model from SynthSeg~\cite{billot2023synthseg} to produce synthetic brain MRIs specifically tailored for the downstream task of brain age prediction. Subsequently, we utilize this generative model to sample MRI scans with random contrasts and resolutions from our cohort of healthy subjects, and we train our brain age model on these synthetic images. The high variability provided by the generative model enables SynthBA to operate with different MRI sequences and resolutions without the need for re-training or fine-tuning while maintaining reasonable accuracy. We evaluate SynthBA using an external dataset comprising MRIs from three distinct MRI sequences at varying resolutions, demonstrating its robustness across various acquisition settings compared to state-of-the-art brain age models. Ultimately, we analyse how the brain PAD obtained using SynthBA from a cohort of subjects with AD correlates with measures of cognitive dysfunction.

\section{Method}
This section describes our methodology, which involves: (i) two adjustments applied to the generative model introduced in SynthSeg~\cite{billot2023synthseg}, and (ii) the development of a neural network to estimate brain age, trained on synthetic brain MRIs produced by the modified generative model. Additionally, we describe the preprocessing steps applied to the brain MRIs used in our study.

\subsection{Adjustments to the generative model from SynthSeg.}

The generative model presented in SynthSeg~\cite{billot2023synthseg}, referred to as $\mathcal{G}$, consists of a series of steps aimed at generating synthetic brain MRIs from brain tissue segmentations. This process is illustrated in Figure~\ref{fig:generator}. A brain tissue segmentation, denoted as $s$ (input A in Figure~\ref{fig:generator}), is a 3D map that assigns each voxel at coordinates $(i, j, k)$ a label $s_{ijk} = l$, identifying a specific brain region. \newline

Starting with a tissue segmentation $s$, the generative process applies a random spatial transformation $\phi$ to produce a deformed segmentation $s' = \phi(s)$ (step B from Figure~\ref{fig:generator}). Originally, in SynthSeg~\cite{billot2023synthseg}, $\phi$ includes both affine transformations and elastic deformations. The \textbf{first adjustment} we applied to the generative model is to restrict the spatial deformation to only a random rigid transformation, explicitly excluding shearing, scaling, and non-linear components. In particular, the exclusion of scaling and elastic deformations aims to prevent transformations that could mimic atrophy patterns associated with ageing, thus avoiding a potential mismatch between the subject's chronological age and the appearance of the transformed segmentation $s'$. \newline

The next step involves generating a 3D image $x$ from the labels $s'$ using a Gaussian Mixture Model (GMM). The GMM assigns a Gaussian distribution $\mathcal{N}(\mu_l, \sigma_l)$ to each segmentation label $l$ indicating a different brain region. The parameters $(\mu_l, \sigma_l)$ are randomly selected either from a fixed uniform prior distribution (option a) or an estimated Gaussian prior distribution (option b): 

$$
\begin{matrix}
\text{(option a)} \\
\mu_l \sim U[\mu_a,\mu_b] \\
\sigma_l \sim U[\sigma_a,\sigma_b]
\end{matrix}
\hspace{1cm}
\begin{matrix}
\text{(option b)} \\
\mu_l \sim \mathcal{N}(\mu_{\mu,l}^k,\sigma_{\mu,l}^k) \\
\sigma_l \sim \mathcal{N}(\mu_{\sigma,l}^k,\mu_{\sigma,l}^k)
\end{matrix}
$$

In SynthSeg~\cite{billot2023synthseg}, the authors provide a method to estimate the prior parameters in (option b) using real images from a specific MRI sequence. Our \textbf{second adjustment} involves expanding this feature to enable the estimation of $K$ sets of prior parameters from $K$ different MRI sequences.  Specifically, the prior parameters $(\mu_{\mu,l}^k, \sigma_{\mu,l}^k, \mu_{\sigma, l}^k, \sigma_{\sigma,l}^k)$ for each label $l$ and sequence $k = 1, \dots, K$ are estimated from brain MRIs reflecting the $k$-th MRI sequence. During generation, one set of parameters is randomly selected from the $K$ available sets. As illustrated in Figure~\ref{fig:generator}, utilizing the uniform prior (option a) results in unrealistic random contrasts (upper branch from Figure~\ref{fig:generator}). Conversely, employing the estimated Gaussian priors (option b) constrains the training domain, yielding more realistic, albeit randomized, MRI contrasts (lower branch from Figure~\ref{fig:generator}). Finally, to generate the synthetic image $x$, each voxel $x_{ijk}$ with label $s_{ijk} = l$ is sampled from $x_{ijk} \sim \mathcal{N}(\mu_l, \sigma_l)$ (step C in Figure~\ref{fig:generator}). \newline

The generative process continues as detailed in SynthSeg~\cite{billot2023synthseg}. Specifically, we follow the implementation in~\cite{ravi2024efficient} to generate a random bias field, which we then apply to the synthetic image $x$ to simulate the intensity inhomogeneity artefact (step D in Figure~\ref{fig:generator}). The image intensities are then rescaled to the range $[0,1]$, and a random gamma transform is applied to further enhance the heterogeneity of the contrast (step E in Figure~\ref{fig:generator}). Finally, the 3D image $x$ is resampled to a random lower resolution, which may be isotropic or anisotropic (step F in Figure~\ref{fig:generator}). Further technical details of each generation step and the parameter ranges for the random transformations are available in \cite{billot2023synthseg}. We use the notation $x \sim \mathcal{G}(s)$ to summarize the process of generating a synthetic scan $x$ from a brain tissue segmentation $s$.

\subsection{Training the Proposed Brain Age Model}
Let $s$ be a brain tissue segmentation from an MRI of a healthy individual. Using the modified generative model, we sample a synthetic brain MRI $x \sim \mathcal{G}(s)$. As mentioned in Section~\ref{sec:introduction} the brain age $y_b$ and the chronological age $y_c$ are equivalent ($y_b = y_c$) for healthy subjects. Therefore, we train a neural network $f_\theta$, with parameters $\theta$, to predict the brain age $f_\theta(x) \approx y_b$ from the synthetic brain MRI $x$. Specifically, our training process aims to find the parameters $\theta$ that minimize the $L_1$ loss $| y_c - f_\theta(x)|$, which is equal to minimizing $| y_b - f_\theta(x)|$. The training pipeline is summarized in Figure~\ref{fig:training-pipeline}. We employ a 3D DenseNet~\cite{huang2017densely} as our parametric model. As the contrast of the synthetic brain MRI is randomly determined at each training iteration, we expect the neural network to estimate the brain age mainly from the structural features of the brain, potentially developing contrast-agnostic properties.

\begin{figure}[h]
    \centering
    \includegraphics[width=\columnwidth]{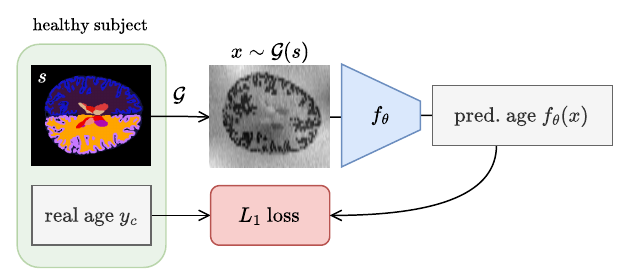}
    \caption{SynthBA training pipeline.}
    \label{fig:training-pipeline}
\end{figure}

\subsection{MRI Preprocessing}
Before training, we aligned all brain MRIs to the standard $1 \text{mm}^3$ MNI space to remove variability arising from different MRI orientations. Subsequently, we used SynthSeg v.2~\cite{billot2023synthseg} to extract brain tissue segmentations from the brain MRIs, to be used within the generative model to sample synthetic scans. The segmentation identifies 32 different brain regions, plus the background. During the training phase, the synthetic scans are downsampled to a resolution of $1.4 \text{mm}^3$, resized to a dimension of $130 \times 130 \times 130$ through padding or cropping, and the intensities are rescaled in the range $[0,1]$.

\section{Experiments and Results}
In this section, we first describe the datasets and training settings used in this study. We then compare SynthBA to state-of-the-art brain age models under various acquisition settings. Finally, we provide a brief clinical evaluation of the brain PAD calculated using SynthBA.

\begin{table}[h]
    \centering
    \caption{Number of healthy subjects and scans in each internal (INT) and external (EXT) dataset for each MRI sequence and their resolutions. Resolutions (R) are classified as Research Grade (RG) or Clinical Grade (CG).}
    \def\arraystretch{1.4}
    \resizebox{\columnwidth}{!}{
    \begin{tabular}{c|c|c|cc|cc|cc}
         \hline
         \multirow{2}{*}{Dataset} & \multirow{2}{*}{Type} & \multirow{2}{*}{\#subj.} & \multicolumn{2}{|c|}{T1w} & \multicolumn{2}{|c|}{T2w} & \multicolumn{2}{|c}{FLAIR}  \\
         \cline{4-9}
          &  & & \#scans & R. & \#scans & R. & \#scans & R.  \\
         \hline
         ADNI    & INT & 784  & 3345 & RG & -    & - & - & - \\
         AIBL    & INT & 192  & 596  & RG & -    & - & - & - \\
         IXI     & INT & 563  & 563  & RG & 584  & RG & - & - \\
         HCP     & INT & 1105 & 1105 & RG & 1105 & RG & - & - \\
         CoRR    & INT & 1461 & 1461 & RG & -    & - & - & - \\
         \hline
         OASIS 3 & EXT & 811 & 1241 & RG & 2389 & CG & 921 & CG \\
         \hline
    \end{tabular}
    }
    \label{tab:datasets}
\end{table}

\begin{table*}[t!]
    \caption{Comparison with state-of-the-art brain age models on internal and external test sets.}
    \centering
    \def\arraystretch{1.4}
    \resizebox{\textwidth}{!}{
    \begin{tabular}{c|l|cc|ccc|c|c}
    \hline
        \textbf{Method} & \textbf{Config} & \textbf{T1w-INT-RG} & \textbf{T2w-INT-RG} & \textbf{T1w-EXT-RG} & \textbf{T2w-EXT-CG} & \textbf{FLR-EXT-CG} & \textbf{AVG-ALL} & \textbf{AVG-EXT} \\ \hline
        \multirow{7}{*}{\cite{wood2024optimising}} 
        & Pre-trained on GRE   & 22.7 ± 8.97 & 11.9 ± 6.92 & 20.4 ± 8.07 & 20.5 ± 8.25 & 21.6 ± 8.46 & 19.4 ± 4.30 & 20.8 ± 0.63\\
        & Pre-trained on T1w   &  13.9 ± 7.71 & 22.5 ± 9.26 &	11.5 ± 6.28	& 13.3 ± 8.19 & 28.4 ± 7.31 & 17.9 ± 7.24 & 17.7 ± 9.28  \\ 
        & Pre-trained on T2w   &  19.7 ± 11.27 & 4.6 ± 4.72 & 18.6 ± 7.07 & 12.2 ± 5.80 & 18.5 ± 8.60 & 14.7 ± 6.35 & 16.4 ± 3.68  \\ 
        & Pre-trained on FLAIR &  11.2 ± 6.09 & 6.9 ± 6.68 & 8.1 ± 5.09 & 21.7 ± 9.09 & 15.9 ± 5.75 & 12.8 ± 6.07 & 15.2 ± 6.83 \\ 
        & Pre-trained on DWI   &  13.5 ± 8.26 & 10.5 ± 7.44 & 13.0 ± 6.76 & 27.0 ± 8.96 & 11.0 ± 5.58 & 15.0 ± 6.84 & 17.0 ± 8.75 \\
        \cline{2-9}
        & Fine-tuned on T1w    & 3.7 ± 3.23 & 7.4 ± 7.73 & 4.3 ± 3.78 & 30.8 ± 9.80 & 7.5 ± 5.05 & 10.7 ± 11.33 & 14.2 ± 14.45 \\ 
        & Fine-tuned on T2w    & 27.9 ± 14.81 & \textbf{3.1 ± 2.54} & 29.2 ± 6.87 & \textbf{5.2 ± 4.29} & 15.8 ± 9.46 & 16.2 ± 12.26 & 16.7 ± 12.07 \\ \hline
        \multirow{2}{*}{\cite{barbano2023contrastive}} 
        & Trained on T1w & \textbf{3.2 ± 2.75} & 11.9 ± 10.89 & \textbf{4.1 ± 3.60} & 11.3 ± 9.75 & 5.9 ± 4.21 & 7.3 ± 4.06 & 7.1 ± 3.76 \\ 
        & Trained on T2w & 16.5 ± 11.00 & 3.4 ± 3.13 & 13.0 ± 8.46 & 7.1 ± 5.80 & 15.3 ± 8.09 & 11.0 ± 5.61 & 11.8 ± 4.24 \\ \hline
        \multirow{2}{*}{Ours} 
        & SynthBA-u & 4.5 ± 4.41 & 4.7 ± 4.90 & 5.0 ± 3.90 & 5.9 ± 5.01 & 6.3 ± 5.38 & 5.3 ± 0.81 & 5.7 ± 0.70 \\
        & SynthBA-g & 3.8 ± 3.78 & 5.1 ± 4.54 & 4.5 ± 4.09 & \textbf{5.2 ± 4.70} & \textbf{5.0 ± 4.75} & \textbf{4.7 ± 0.61} & \textbf{4.9 ± 0.40} \\ \hline
    \end{tabular}}
    \label{tab:comparison}
\end{table*}

\subsection{Datasets}
\label{sec:datasets}
We collected T1w, T2w, and FLAIR brain MRIs from healthy subjects using six publicly available clinical datasets: ADNI~\cite{petersen2010alzheimer}, AIBL~\cite{ellis2009australian}, OASIS 3~\cite{lamontagne2019oasis}, IXI~\cite{ixi}, HCP~\cite{van2013wu}, and CoRR~\cite{zuo2014open}. Table~\ref{tab:datasets} reports the number of scans in each dataset for each MRI sequence and their resolutions. We define a research-grade MRI as an isotropic scan with a voxel resolution of approximately $1\text{mm}^3$ or finer. In contrast, a clinical-grade MRI is defined as an anisotropic scan or a low-resolution isotropic MRI. We aggregate five datasets (ADNI, AIBL, IXI, HCP, CoRR) into a combined internal dataset comprising 4,105 healthy subjects. This dataset includes 7,060 research-grade T1w MRIs and 1,689 research-grade T2w MRIs. The subjects' ages range from 6 to 95 years, displaying a bimodal distribution with peaks at 25 and 74 years. Females constitute 53\% of the subjects. We divided the subjects into training, validation, and test sets in proportions of 85\%, 5\%, and 10\%, respectively. We refer to the T1w and T2w research-grade MRIs from the internal test set as \textbf{T1w-INT-RG} and \textbf{T2w-INT-RG}, respectively. To evaluate the generalizability of the model, we use OASIS 3 as the external dataset, ensuring that no MRIs from this dataset were used during training or for model selection. The average age in the OASIS 3 dataset is $68 \pm 9$ and 61\% of the subjects are female. We divided OASIS 3 into three separate test sets: research-grade T1w images (\textbf{T1w-EXT-RG}), clinical-grade T2w images (\textbf{T2w-EXT-CG}), and clinical-grade FLAIR images (\textbf{FLR-EXT-CG}). The lowest resolution encountered in clinical-grade MRIs was $6 \text{mm}$ along the acquisition direction. Finally, we selected 675 AD subjects from ADNI, considering for each subject only one research-grade T1w MRI and the associated ADAS-Cog-13 cognitive score. We refer to this dataset as \textbf{AD-EVAL}.

\subsection{Experiments settings}
We explore two different configurations of SynthBA, depending on which type of prior is used in the GMM step within the generative model. The first configuration, named \textbf{SynthBA-u}, utilizes a fixed uniform prior (refer to the upper branch in Figure~\ref{fig:generator}). The second configuration, referred to as \textbf{SynthBA-g}, incorporates two Gaussian priors (refer to the lower branch in Figure~\ref{fig:generator}), whose parameters are estimated from T1w and T2w training scans, respectively. The brain tissue segmentations, derived from the internal training set of T1w images, are utilized by the generative model to dynamically produce synthetic brain MRIs during training. We train SynthBA on these synthetic brain MRIs for 300 epochs. We use the AdamW optimizer~\cite{loshchilov2017decoupled} with a batch size of 16 and an initial learning rate set to $10^{-4}$, which is progressively reduced to $10^{-5}$ using a cosine decay scheduler~\cite{loshchilov2016sgdr}. We use a dropout probability of 30\%. To mitigate the non-uniform age distribution in the training set, we divide the age range into five-year bins and allocate each training sample to its corresponding bin. During training, we employ uniform sampling across these bins to create a batch, effectively balancing the inclusion of all age groups. We use the validation set to select the model with the best performance. For the 3D DenseNet, we use the implementation available in MONAI~\cite{cardoso2022monai}. All experiments are conducted on a GeForce RTX 4090 (24GB VRAM).

\subsection{Comparative study}
We conduct an evaluation of both SynthBA-g and SynthBA-u configurations, on the five different test sets detailed in Section~\ref{sec:datasets} (two internal, three external), with varying MRI sequences and resolutions. We compare our models with state-of-the-art baselines: \cite{wood2024optimising} and \cite{barbano2023contrastive}. For the first baseline~\cite{wood2024optimising}, we evaluate (i) their five publicly-available pre-trained brain age models, each one specialized on a distinct MRI sequence (GRE, DWI, T1w, T2w, FLAIR), and (ii) their T1w and T2w models fine-tuned respectively on our T1w and T2w training images. For the second baseline~\cite{barbano2023contrastive}, there are no publicly-available models; therefore, we evaluate a T1w model and a T2w model trained with their contrastive method on our T1w and T2w training images, respectively. Table~\ref{tab:comparison} presents the obtained results. Here we report the Mean Absolute Error (MAE) for each method across all internal and external test sets. Importantly, we report the average MAE over all test sets (AVG-ALL) and the average MAE over external test sets only (AVG-EXT). Both average MAEs serve as key metrics for this study, summarizing brain age model performance across various MRI sequences and resolutions. \newline

Table~\ref{tab:comparison} reveals that some baseline models achieve considerably low MAE on specific MRI sequences and resolutions compared to SynthBA. However, these models fail to generalize across heterogeneous acquisition settings, with their AVG-ALL and AVG-EXT errors ranging from 7 to 19 years and 7 to 20 years, respectively. In contrast, both SynthBA-u and SynthBA-g demonstrate robust performance across various acquisition settings without requiring re-training. In particular, SynthBA-g achieves the lowest average MAEs of $4.7 \pm 0.61$ and $4.9 \pm 0.40$ for AVG-ALL and AVG-EXT, respectively. Moreover, SynthBA-g yields the best MAE ($5.0 \pm 4.75$) on FLR-EXT-CG and matches the best MAE ($5.2 \pm 4.70$) on T2w-EXT-CG. We observe that SynthBA-g consistently achieves a lower MAE compared to SynthBA-u and demonstrates superior generalization capability when applied to external datasets. This could be attributed to the fact that the synthetic MRIs generated with estimated Gaussian priors from two different MRI sequences (T1w, T2w) more closely resemble real MRIs.

\subsection{Clinical evaluation}
In this experiment, we conduct a clinical evaluation of SynthBA, examining the relationship between the ADAS-Cog13 score and brain PAD determined by the SynthBA model. The ADAS-Cog13~\cite{mohs1997development} score serves as a cognitive evaluation tool specifically designed for assessing cognitive function in AD patients. Using the best SynthBA configuration (SynthBA-g), we extract predicted brain age data from subjects within the AD-EVAL dataset and compute the brain PAD. Our analysis reveals a significant Pearson correlation of 0.275 ($p<0.001$) between brain PAD and the ADAS-Cog-13 score among AD subjects, as depicted in Figure~\ref{fig:adascogreg}. 

\begin{figure}[h]
    \centering
    \includegraphics[width=\columnwidth]{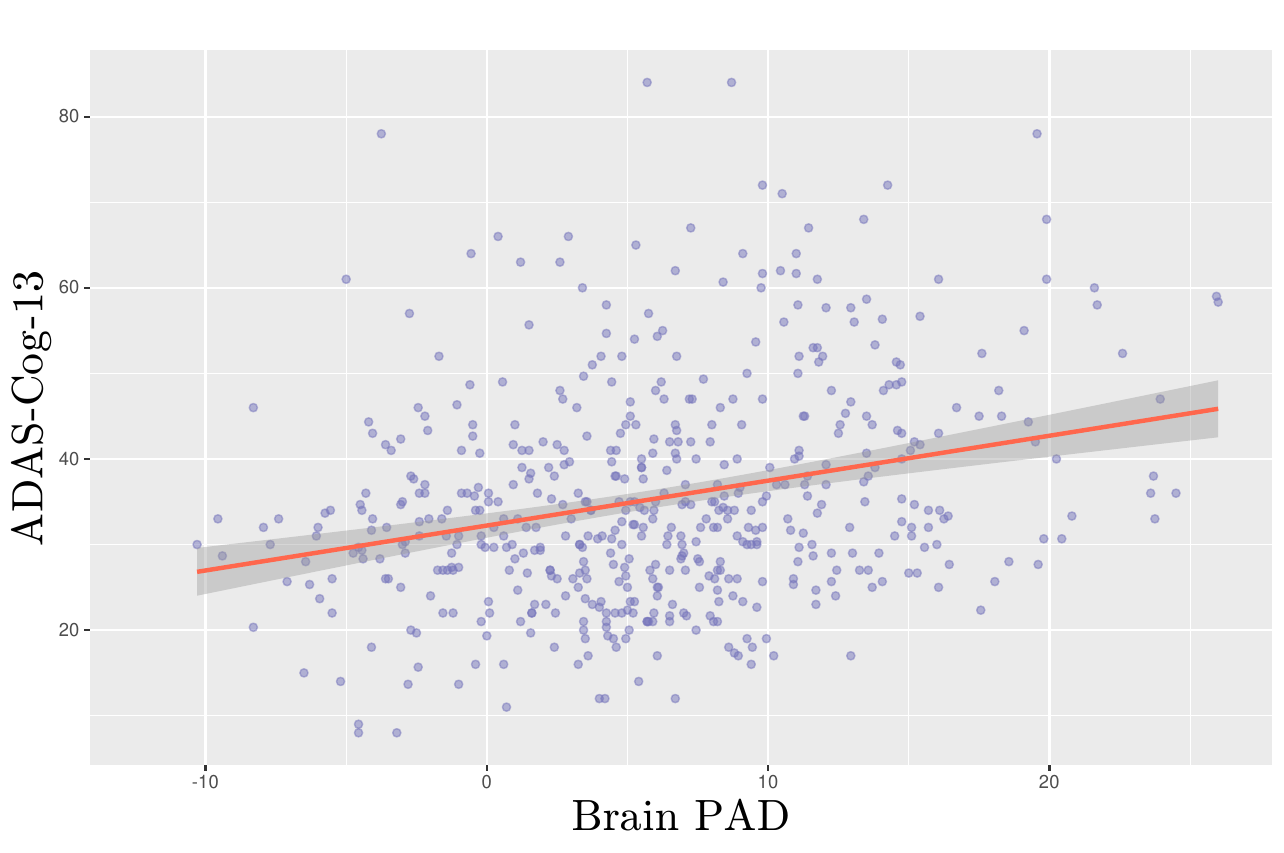}
    \caption{Association between SynthBA's brain PAD and ADAS-Cog13.}
    \label{fig:adascogreg}
\end{figure}

\section{Conclusions}
In this study, we introduce SynthBA, a robust brain age model that maintains consistent performance across different acquisition settings. We demonstrate its capabilities on both research-grade and clinical-grade MRIs from three different sequences (T1w, T2w, and FLAIR) and compare its performance to state-of-the-art brain age models, showcasing its superiority in heterogeneous acquisition settings. Future directions include enhancing the generative model with various MRI artefacts to increase predictive robustness and incorporating random orientation to avoid alignment to a common template, especially in low-resolution, anisotropic scans, where alignment can be detrimental. Our work encourages the widespread adoption of brain age analysis by extending its application to scenarios where retraining or fine-tuning is impractical. To facilitate this, we publicly release our SynthBA models as a standalone software package\footnote{https://github.com/LemuelPuglisi/SynthBA}.

\section*{Acknowledgments}

Lemuel Puglisi is enrolled in a PhD program at the University of Catania, fully funded by PNRR (DM 118/2023). Francesco Guarnera is funded by the PNRR MUR project PE0000013-FAIR. Alessia Rondinella is a PhD candidate enrolled in the National PhD in Artificial Intelligence, XXXVII cycle, course on Health and life sciences, organized by Università Campus Bio-Medico di Roma.
\balance
\bibliographystyle{IEEEtran}
\bibliography{IEEEabrv,bibliography}

\end{document}